# Vertically Correlated Disorder and Structured Interlayer Tunneling in Cuprates

E.Yu. Beliayev[1,2*], Y.K. Mishra[2], I.G. Mirzoiev[1], V.V. Andrievskii[1], A.V. Terekhov[1]

[1]B. Verkin Institute for Low Temperature Physics and Engineering of the National Academy of Sciences of Ukraine.

[2]Smart Materials Group, Mads Clausen Institute, University of Southern Denmark, Sønderborg.

e-mail*: *beliayev@ilt.kharkov.ua*

## Abstract

Cuprate superconductors display robust in-plane electronic correlations but exceptionally fragile interlayer coherence. We suggest that even weak vertically correlated disorder (arising from interstitial-oxygen staging, twin boundaries, extended strain fields, or defect-pinned charge textures) can impose a layer-dependent modulation of the interlayer tunneling amplitude $t_\perp(z)$. Because the bare interlayer coupling is intrinsically small, such modulations generate an effectively multichannel $c$-axis electrodynamic response, consistent with multi-component Josephson plasma resonances, nonmonotonic $c$-axis resistivity, redistribution of bilayer magnetic spectral weight, and field-enhanced vertical CDW correlations. We propose a phenomenological framework in which the organization of disorder, rather than its magnitude, governs the effective interlayer coupling and its electrodynamic signatures. This viewpoint unifies diverse $c$-axis anomalies across several cuprate families, suggesting that controlling vertical disorder correlations offers a viable pathway for tuning dimensionality and interlayer coherence in high-$T_c$ superconductors.

## I. Introduction

Layered cuprate superconductors are composed of electronically active $CuO_2$ planes separated by comparatively inert spacer layers. This structural motif produces an extreme electronic anisotropy: while quasiparticle states within the $CuO_2$ planes remain robust and strongly correlated, coherence along the crystallographic $c$ axis is notably fragile. This fragility manifests in a broad range of experimental observations, including the Josephson plasma resonance (JPR), unconventional $c$-axis transport, field-induced three-dimensional charge-density-wave (CDW) correlations, and the sensitivity of bilayer magnetic excitations to dilute planar impurities. Despite extensive study, the microscopic origin of this fragility and the substantial variability in $c$-axis properties across samples and materials remain incompletely understood.

A further complication is that experimentally observed $c$-axis anomalies do not fit naturally into a single classification scheme. Some cuprates exhibit multi-component or broadened JPR spectra, implying significant spatial variations in interlayer tunneling [1], whereas others display a single sharp resonance. In certain compounds, $\rho_c(T)$ follows variable-range-hopping behavior characteristic of strongly localized out-of-plane transport, while in others it evolves through non-monotonic crossovers. Magnetic fields can drive underdoped $YBa_2Cu_3O_{6+\delta}$ into a three-dimensionally coupled CDW state [2], but Hg-based cuprates (among the structurally cleanest) support only short-range, essentially two-dimensional CDW correlations even under comparable conditions [3]. In addition, very small concentrations of planar impurities such as Zn or Ni can substantially modify the bilayer magnetic resonance in $YBa_2Cu_3O_{6+x}$, altering its energy, spectral weight, and line shape while leaving in-plane superconductivity largely intact [4] [5]. These diverse behaviors are often interpreted within separate frameworks, which obscures the broader organizing principles governing $c$-axis electrodynamics. A complementary transport-based perspective on disorder-driven crossovers and the emergence of coherence in low-doped $La_{2-x}Sr_xCuO_{4+\delta}$ was developed in [6], where nonlinear transport regimes and percolative superconducting connectivity were analyzed in detail.

In this work, we argue that many of these apparently unrelated phenomena can be understood within a common framework by recognizing the role of vertically correlated disorder — forms of quenched disorder whose spatial organization extends coherently along the $c$ axis. Although disorder in cuprates is frequently treated as a random, pair-breaking perturbation, structural studies demonstrate that interstitial oxygen and related defects can form vertically extended configurations persisting across multiple unit cells [7]. At the same time, analyses of fluctuating stripe-like and nematic electronic textures indicate that correlated spatial organization of this kind can arise naturally from electronically driven self-organization [8]. The interlayer tunneling amplitude $t_\perp$ is intrinsically small and strongly dependent on local structural geometry, including the apical-oxygen position, the Cu–O–Cu bond configuration, and nematic distortions of the underlying electronic dispersion [9]. The combination of weak interlayer coupling and vertically organized disorder, therefore, produces a layer-dependent tunneling landscape that can significantly affect $c$-axis electrodynamics, even in the absence of charge transfer or genuine three-dimensional electronic order.

The central idea of this work is that the *spatial organization* of disorder, rather than its overall magnitude, can play a decisive role in shaping interlayer coupling in cuprates. To explore this notion, we introduce a minimal phenomenological description in which the effective tunneling amplitude varies from layer to layer as

$$t_\perp(z) = t_0 + \delta t(z),$$

with $\delta t(z)$ containing correlations that extend along the crystallographic $c$ axis. In this picture, different vertically extended regions act as parallel tunneling channels, each contributing a distinct electrodynamic response. Such a multichannel structure naturally accommodates a variety of observed behaviors, including multi-component JPR spectra, sample-dependent forms of $\rho_c(T)$, field-enhanced CDW stacking in $YBa_2Cu_3O_{6+\delta}$, and impurity-induced modifications of the bilayer magnetic resonance.

Our goal is not to construct a microscopic model for $t_\perp(z)$, but to identify the essential consequences of vertically organized disorder for interlayer electrodynamics. This framework provides a common language for discussing several experimental trends and may offer useful guidance for approaches involving controlled disorder, oxygen-order manipulation, or strain-based tuning of interlayer coherence.

The remainder of this article is organized as follows. Section II reviews the fragility of interlayer coherence arising from in-plane inhomogeneity. Section III summarizes structural motifs that can generate vertically correlated disorder. Section IV introduces the phenomenological description of a modulated $t_\perp(z)$. Section V discusses experimental signatures consistent with this approach. Section VI considers implications for competing orders and dimensional crossover. Section VII concludes with a brief summary and outlook.

## II. Fragility of Interlayer Coherence in the Presence of In-Plane Inhomogeneity

Interlayer electronic coupling in cuprate superconductors is intrinsically weak, strongly anisotropic, and extraordinarily sensitive to local structural conditions. Unlike conventional layered metals, cuprates exhibit no coherent $c$-axis band dispersion; out-of-plane transport instead proceeds via tunneling across spacer layers that contain apical oxygens, distorted $CuO_6$ octahedra, and charge-reservoir blocks. Consequently, even subtle variations in Cu–O–Cu bond angles, apical-oxygen positions, or local strain can induce disproportionately large changes in the interlayer tunneling amplitude.

This intrinsic sensitivity is further amplified by the pronounced electronic inhomogeneity within the $CuO_2$ planes. A similar susceptibility to spatial variations appears in the interlayer components of collective electronic orders. Experiments indicate that each plane comprises nanoscale regions with distinct superconducting gaps, carrier densities, and ordering tendencies, and these features rarely align laterally from one layer to the next. A metallic region in one $CuO_2$ plane may therefore overlap with a pseudogapped or insulating region in an adjacent plane, suppressing the coherence of the interlayer tunneling channel connecting them. As a result, *c*-axis transport can exhibit localization- or VRH behavior, even while in-plane transport remains comparatively coherent, as observed in underdoped $La_{2-x}Sr_xCuO_4$ [10]. In lightly doped $La_2CuO_{4+\delta}$, detailed transport studies have demonstrated that the variable-range-hopping regime exhibits a three-dimensional character, despite the strongly layered crystal structure [11]. This observation implies that hole motion can involve interlayer pathways mediated by apical oxygen and defect configurations, reinforcing the notion that even nominally two-dimensional cuprates may sustain vertically extended transport networks under appropriate disorder conditions. Additional evidence for strong in-plane electronic inhomogeneity, capable of frustrating interlayer coherence, is provided by transport measurements in lightly doped LSCO that reveal self-organized anisotropic electronic textures [12].

Interlayer components of collective electronic orders exhibit a comparable fragility. In $La_{2-x}Ba_xCuO_4$ near the 1/8 anomaly, the 90° rotation of stripe orientation between adjacent $CuO_2$ layers frustrates interlayer Josephson coupling and leads to predominantly two-dimensional superconducting behavior with negligible *c*-axis coherence [13] [14] [15]. In $YBa_2Cu_3O_{6+\delta}$, charge-density-wave correlations remain essentially two-dimensional at zero magnetic field and develop genuine three-dimensional stacking only under strong fields that align otherwise misregistered CDW domains [16]. In bilayer cuprates, even dilute planar impurities, such as Zn or Ni, can strongly modify the magnetic resonance, collapsing the odd–even bilayer splitting and thereby revealing the exceptional sensitivity of interlayer magnetic coherence to disorder [4] [5].

Taken together, these observations emphasize a central principle: interlayer coherence is among the most fragile elements of the cuprates' correlated electronic structure. Its pronounced sensitivity to in-plane inhomogeneity, disorder, and local structural distortions makes the *c*-axis response a particularly informative probe of nanoscale electronic organization. At the same time, this susceptibility suggests that disorder, when spatially correlated rather than random, can reorganize interlayer coupling in structured and reproducible ways, enabling the emergent *c*-axis phenomena examined in the sections that follow.

### III. Vertically Correlated Disorder: Microscopic Mechanisms

A wide variety of structural motifs in cuprate superconductors gives rise to disorder whose spatial distribution is not fully random but exhibits correlations extending along the crystallographic *c*-axis. Because the interlayer tunneling amplitude $t_\perp$ is extremely sensitive to apical-oxygen height, Cu–O coordination geometry, and local octahedral tilts, vertically correlated disorder naturally imprints a structured modulation onto the effective interlayer coupling. Several experimentally established mechanisms can produce such vertical organization.

One prominent example arises from interstitial-oxygen staging in $La_2CuO_{4+\delta}$. Neutron and X-ray scattering studies have shown that excess oxygen does not occupy interstitial sites randomly, but forms periodically repeated oxygen-rich layers or domains with a characteristic *c*-axis periodicity [7] [17]. These vertically extended configurations distort the $CuO_6$ octahedra across several adjacent planes, inducing correlated modulations in apical-oxygen displacement and octahedral tilt patterns and thereby generating corresponding variations in $t_\perp(z)$.

Twin boundaries provide a second mechanism for vertically correlated structural disorder. In orthorhombic cuprates such as $YBa_2Cu_3O_{7-\delta}$, twin boundaries form during the tetragonal–orthorhombic transition and propagate as extended planar defects through large portions of the crystal. Low-temperature STM measurements directly reveal densely spaced, quasi-periodic streaks on cleaved $YBa_2Cu_3O_7$ surfaces, consistent with twin boundaries extending over hundreds of nanometers [18]. These defects generate long-range strain fields that alter local Cu–O coordination and thus can influence apical-oxygen displacement and interlayer tunneling pathways. As a result, twin boundaries act as quasi-one-dimensional channels of correlated distortion, structuring the local tunneling environment far more coherently than isolated point defects.

A third class of vertically correlated disorder originates from dislocation lines and extended defect filaments. High-resolution STEM studies of $YBa_2Cu_3O_{7-\delta}$ nanocomposites have revealed that intergrowths of Cu–O chain layers, threading dislocations, and associated three-dimensional strain fields extend over several consecutive unit cells along the crystallographic $c$ axis, forming quasi-one-dimensional columns of correlated structural distortion [19]. These extended defects strongly perturb the local Cu–O coordination environment, bending $CuO_2$ planes and modulating apical-oxygen positions, thereby generating a vertically structured pattern of interlayer tunneling amplitudes. Complementary NQR studies on $La_{2-x}Sr_xCuO_4$ further indicate that dopants can self-organize into nanoscale segregated domains, acting as electronically distinct regions within each $CuO_2$ plane, and providing an additional electronic pathway for layer-by-layer inhomogeneity [20]. When such structural and electronic inhomogeneities persist coherently across successive layers, they form vertically correlated impurity environments capable of modulating $t_\perp$ in a block-like, segmental manner.

Beyond structural defects such as interstitial-oxygen staging and twin boundaries, vertically correlated disorder may also arise from electronically driven ordering tendencies. Recent cross-sectional STM/STS measurements on $YBa_2Cu_3O_{6.81}$ have revealed nanoscale charge-order domains that are correlated across adjacent CuO-chain and $CuO_2$-plane layers, displaying a robust anti-phase alignment of CDW maxima along the $c$ axis [21].

Short-range charge-density-wave textures provide an additional mechanism through which vertical correlations may emerge. In underdoped cuprates, CDW correlations are predominantly two-dimensional, with short in-plane correlation lengths and negligible intrinsic coherence along the $c$ axis [22] [23]. However, local disorder frequently pins these CDW modulations into nanoscale domains whose internal structure remains topologically rigid [24]. When similar pinning environments recur across successive layers (whether due to repeated defect arrays, vertically extended strain fields, or correlated dopant configurations), CDW domains can acquire weak but finite interlayer alignment even in the absence of genuine long-range three-dimensional order. Magnetic fields further enhance such tendencies by suppressing superconducting fluctuations and stabilizing interlayer CDW stacking, as demonstrated in $YBa_2Cu_3O_{6+\delta}$ under high magnetic fields [16].

Nonlinear lattice dynamics may furnish an additional, time-dependent route to vertical correlation. Strongly anharmonic oxides can support intrinsic localized modes (ILMs) — large-amplitude, spatially confined vibrational excitations that arise from nonlinearities in the lattice potential [25]. If present in cuprates, such excitations could intermittently synchronize distortions of neighboring $CuO_6$ octahedra across multiple layers, generating dynamically correlated fluctuations in the interlayer tunneling amplitude $t_\perp(z)$. Although direct experimental confirmation of breather-like modes in cuprates remains outstanding, their existence in other strongly anharmonic crystals illustrates the broader principle that oxygen lattice dynamics can mediate correlations between structurally distant layers.

Taken together, these mechanisms indicate that disorder in cuprates is often highly organized rather than random: extended defects, dopant clustering, pinned electronic textures, and dynamic lattice excitations all introduce vertical correlations whose influence naturally propagates into the interlayer tunneling landscape $t_\perp(z)$. This vertically structured disorder plays a central role in shaping the emergent *c*-axis electrodynamics examined in the following sections.

### IV. Phenomenological Framework for Vertically Structured Interlayer Tunneling

Because interlayer coupling in cuprate superconductors is intrinsically small, even modest vertically correlated perturbations can produce substantial rearrangements of the *c*-axis electrodynamics. We capture this sensitivity by introducing a phenomenological description in which the interlayer tunneling amplitude becomes explicitly layer dependent,

$$t_\perp \rightarrow t_\perp(z),$$

where $z$ indexes successive $CuO_2$ planes. Correlated disorder (whether of structural, chemical, or electronic origin) modulates the local Cu–O coordination environment and apical-oxygen geometry, creating contiguous vertical segments in which $t_\perp(z)$ is enhanced or suppressed.

To leading order, the tunneling profile may be expressed as

$$t_\perp(z) = t_0 + \delta t(z),$$

where $t_0$ is the average interlayer matrix element and $\delta t(z)$ encodes the spatially correlated disorder. In contrast to conventional incoherent-tunneling approaches [26], here $\delta t(z)$ is not treated as uncorrelated noise but as a quantity with finite-range correlations along the *c*-axis:

$$\langle \delta t(z)\, \delta t(z+n) \rangle \neq 0.$$

Such correlations produce distinct "tunneling channels" — vertical segments in which the effective coupling is locally uniform but varies between segments. Because interlayer coherence in cuprates is already marginal, these variations generate an electrodynamic response that reflects the *distribution* of $t_\perp(z)$, rather than its mean value.

This vertically structured tunneling profile does not imply real charge transfer between layers nor the emergence of genuine three-dimensional electronic order. Instead, it represents a modulation of the interlayer matrix element that governs coupling between superconducting and normal-state degrees of freedom in adjacent $CuO_2$ planes. Since $t_\perp$ directly influences the Josephson plasma resonance frequency [27], the *c*-axis optical conductivity [28], and the interlayer coupling [29], variations in $t_\perp(z)$ naturally lead to a multichannel *c*-axis response. Each vertical segment of similar $t_\perp$ contributes a characteristic energy scale (e.g., a distinct plasma frequency, activation energy, or magnetic resonance component) resulting in the experimentally observed superposition of modes. Such vertically correlated disorder may naturally emerge under conditions where defect generation is itself spatially correlated along the *c*-axis, for example, in strongly irradiated $YBa_2Cu_3O_{7-\delta}$ crystals, where pronounced irradiation-induced anomalies in *c*-axis coherence and fluctuation conductivity regimes have been reported Solovjov et al. [30].

In this picture, differences between nominally similar samples arise not from their average doping or impurity concentration but from the *organization* of disorder that structures the tunneling landscape. Two crystals with the same chemical composition may display dramatically different *c*-axis resistivity,

JPR line shapes, or CDW stacking tendencies if their defect architectures differ in vertical correlation length or amplitude. Similar conclusions have been drawn in studies emphasizing the role of nanoscale inhomogeneity in shaping electronic responses [31].

This framework provides a natural explanation for why samples with similar average doping or impurity concentrations can exhibit markedly different *c*-axis properties: the organization of disorder, rather than its overall magnitude, determines the vertical structure of the tunneling environment. A schematic illustration of the vertically structured tunneling profile and its electrodynamic consequences is shown in **Fig. 1**. The next section surveys the experimental evidence that naturally follows from this perspective.

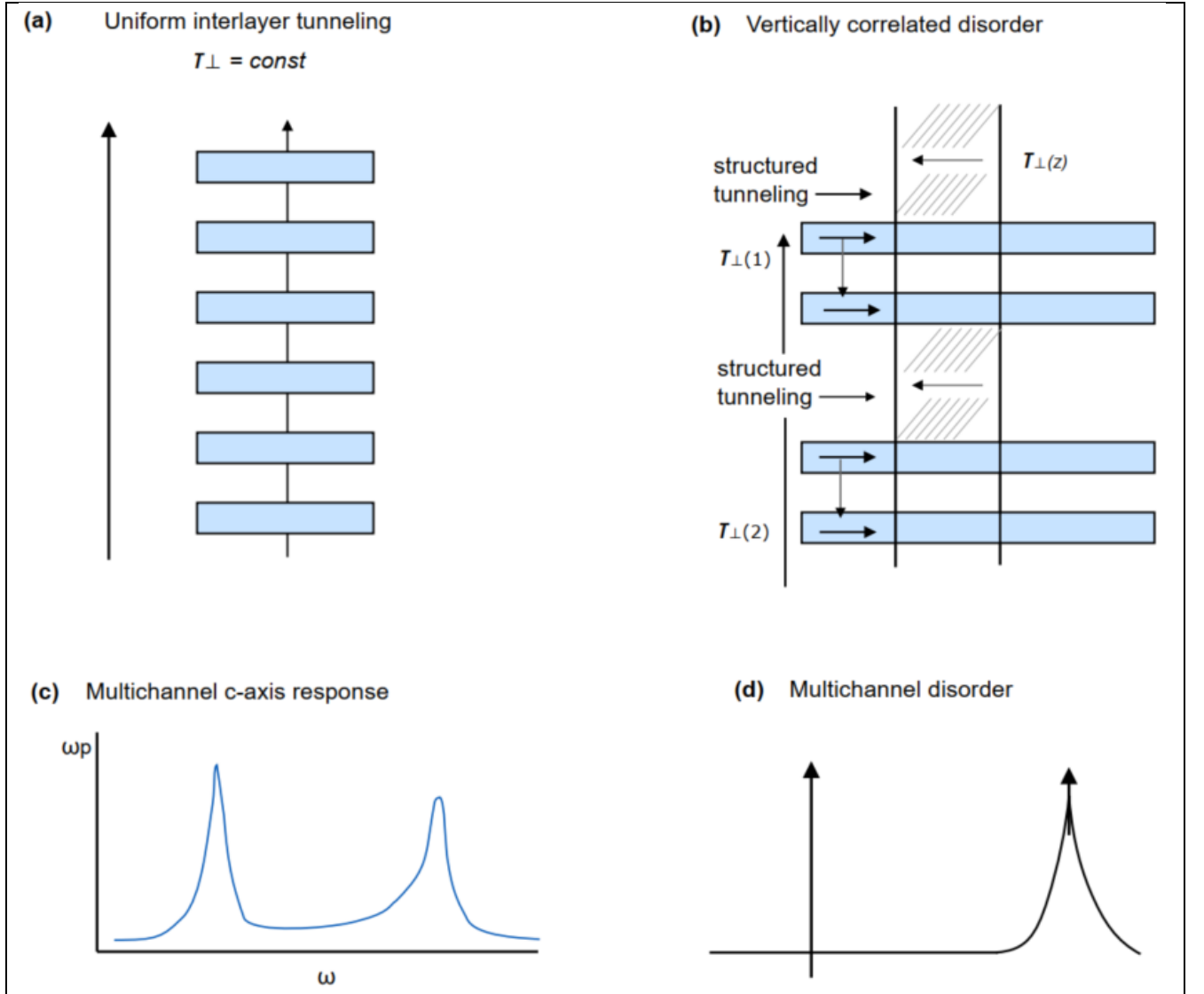

**Fig. 1.** (*a*) *Uniform interlayer tunneling*: a perfectly ordered layered structure with a constant interlayer matrix element $t_\perp = const$; (*b*) *Vertically correlated disorder*: defect strings or extended correlated defects generate extended regions with modified local transparency, producing a depth-dependent tunneling profile $t_\perp(z)$ composed of segments with distinct amplitudes $t_\perp(1)$ and $t_\perp(2)$. Such structured tunneling channels reflect the correlated nature of disorder along the *c*-axis. (*c*) *Multichannel c-axis response*: a depth-dependent tunneling landscape naturally leads to multiple Josephson plasma resonances associated with inequivalent tunneling segments. (*d*) *Multichannel disorder distribution*: schematic representation of the emergent distribution of tunneling strengths arising from vertically correlated defects.

**V. Experimental Signatures of Vertically Structured Interlayer Tunneling**

A broad spectrum of experimental observations in cuprate superconductors, often interpreted through disparate microscopic mechanisms, can be naturally understood once the interlayer tunneling amplitude is allowed to vary spatially due to vertically correlated disorder. Because $t_\perp$ is intrinsically small and highly sensitive to local structural environments, even modest vertical correlations among dopants, strain fields, or oxygen configurations can give rise to distinct tunneling channels. These channels manifest directly in the *c*-axis electrodynamics, providing a unifying interpretation of phenomena involving Josephson plasma resonances, *c*-axis transport, charge-density-wave stacking, bilayer magnetic excitations, and local magnetic or electronic probes.

One of the clearest manifestations of vertically structured interlayer tunneling is the Josephson plasma resonance. In an ideally uniform layered superconductor, the JPR appears as a single, sharply defined mode reflecting a homogeneous interlayer superfluid stiffness [27]. In several cuprate families, however, the JPR deviates substantially from this ideal behavior. In $La_{2-x}Sr_xCuO_4$, the JPR becomes strongly broadened and asymmetric, a hallmark of a broad distribution of local plasma frequencies generated by nanoscale variations of the interlayer Josephson coupling [32].

A closely related source of such non-uniform interlayer coupling arises in oxygen-rich $La_2CuO_{4+\delta}$. Neutron-diffraction studies reveal that the incorporation of excess oxygen drives a reversible phase-separation process, forming vertically extended oxygen-rich and oxygen-poor domains below ~320 K [33]. These domains distort the local Cu–O environment and modulate the interlayer spacing, naturally producing spatial variations in the effective tunneling amplitude $t_\perp$, even though the JPR itself has not been measured in this compound.

Even more pronounced multimodal behavior occurs in systems with inequivalent or spatially non-uniform interlayer junctions. In $Bi_2Sr_2CaCu_2O_{8+\delta}$, for example, spatial variations of the interlayer phase coherence lead to multiple JPR branches and nontrivial field- and frequency-dependent mode evolution [34]. Likewise, nonlinear terahertz spectroscopy reveals distinct Josephson-plasmon branches in layered cuprates, providing direct evidence that non-uniform interlayer coupling gives rise to multiple collective modes even when only a single plasma edge is visible in linear optics [35].

Within the present framework, these broadened or multi-component spectra arise naturally when correlated disorder partitions the crystal into vertically extended segments with different effective tunneling strengths. Segments with enhanced $t_\perp$ contribute higher-frequency components, while regions with suppressed coupling generate lower-frequency modes, so that the measured JPR line shape becomes a superposition of the tunneling channels encoded in $t_\perp(z)$.

An analogous multichannel response emerges in the DC limit: *c*-axis charge transport exhibits a similarly rich phenomenology. Underdoped cuprates often exhibit activated or VRH–like behavior along the *c*-axis [12], despite in-plane transport remaining metallic. In many samples, $\rho_c(T)$ develops nonmonotonic temperature dependence, plateau-like regions, or multiple crossover scales, often differing markedly between crystals of the same nominal composition. Such behavior arises naturally if vertically correlated disorder produces a distribution of tunneling channels that contribute in parallel. Similar transport phenomenology, in which conduction is governed by a limited number of spatially organized tunneling paths rather than by average disorder, has been documented in granular oxide systems exhibiting percolative transport [36]. In such systems, the morphology and spatial correlation of conducting elements determine the effective dimensionality and the emergence of multichannel response. Related behavior has also been observed in structurally inhomogeneous metallic and granular films [37] [38], where spatially heterogeneous tunneling barriers give rise to activated and VRH-like transport regimes. While microscopically distinct from layered cuprates, these systems illustrate a broader principle: transport in strongly anisotropic or heterogeneous media can be dominated by a structured network of preferential tunneling links rather than by the mean defect concentration. In cuprates, modest variations in oxygen ordering, annealing conditions, or local strain can play an analogous role by shifting the relative weights of different $t_\perp(z)$ segments, thereby producing robust sample-specific differences in the $c$-axis resistivity.

Charge-density-wave correlations offer another striking manifestation of vertical structuring. CDW order in underdoped cuprates is primarily two-dimensional [22], yet under strong magnetic fields, it can develop three-dimensional stacking, most prominently in YBa$_2$Cu$_3$O$_{6+\delta}$ [16] [2]. This field-induced stacking should not be interpreted as disorder creating long-range three-dimensional CDW order; rather, vertically correlated disorder can prearrange weak interlayer couplings among short-range CDW domains. Suppressing superconducting fluctuations with a magnetic field reveals and amplifies this latent tendency. The absence of any comparable 3D stacking in structurally cleaner single-layer cuprates such as HgBa$_2$CuO$_{4+\delta}$, where CDW correlations remain short-ranged and essentially two-dimensional [3], underscores that the emergence of 3D order in YBa$_2$Cu$_3$O$_{6+\delta}$ relies sensitively on the underlying organization of disorder. Consistent with this view, NMR studies indicate that in YBa$_2$Cu$_3$O$_y$, short-range CDW modulations in the normal state are pinned and stabilized by native defects rather than merely determined by their overall concentration [39].

Neutron-scattering studies of bilayer cuprates offer a further window into the sensitivity of interlayer correlations to vertically structured disorder. In an ideal CuO$_2$ bilayer, the magnetic resonance splits into odd and even components whose relative intensities depend on the coherence of the interlayer magnetic coupling [40]. This bilayer coherence is surprisingly fragile: even dilute Zn substitution strongly perturbs the local magnetic environment, broadening the resonance in both momentum and energy space, redistributing spectral weight and effectively suppressing the interlayer coherence required for a well-defined odd–even structure [4]. Within the current framework, this behavior follows naturally: vertically correlated disorder acts differently on the two CuO$_2$ planes within a bilayer, disrupting their magnetic equivalence and modulating the effective interlayer matrix element without requiring changes in carrier density or long-range magnetic order.

Local probes provide complementary constraints on the mesoscale organization of electronic and magnetic inhomogeneities in cuprates. μSR and NMR measurements in La-214 compounds reveal slowly fluctuating, spatially heterogeneous magnetic regions whose characteristic length scales exceed those of individual CuO$_2$ planes [41]. Although these techniques do not directly resolve correlations along the $c$ axis, the persistence and mesoscale extent of the magnetic domains imply that they are stabilized by an underlying structural landscape that varies only weakly between adjacent layers. Scanning tunneling spectroscopy on Bi-based cuprates reveals an analogous electronic phenomenology. STM studies detect robust nanoscale variations of the superconducting gap [42] and bond-centered unidirectional electronic domains [43], both of which display remarkable spatial stability and evolve only slowly across the field of view. The longevity and coherence of these textures indicate that the structural or chemical degrees of freedom that template them must themselves be extended over distances larger than a single CuO$_2$ layer.

Direct evidence for such subsurface structural organization is provided by spectroscopic-imaging STM of Bi$_2$Sr$_2$CaCu$_2$O$_{8+\delta}$. Variations of the superconducting gap $\Delta(r)$ follow the phase of the incommensurate bulk supermodulation of the lattice, which periodically distorts local CuO$_5$ pyramids and modulates out-of-plane atomic positions over a characteristic length of several nanometers (≈25–30 Å) [44]. Because this supermodulation is intrinsically three-dimensional and coherent over many successive layers, its imprint on the surface LDOS demonstrates how vertically correlated structural motifs, such as modulated apical-oxygen positions, defect columns, or strain-organized regions, can stabilize the ostensibly two-dimensional gap textures observed by STM. Taken together, these observations suggest that slowly varying, vertically correlated components of the disorder landscape provide a crucial organizing framework for the electronic and magnetic heterogeneity characteristic of underdoped cuprates.

Viewed in this light, the complex interlayer phenomenology of cuprates does not require genuine *c*-axis charge ordering or long-range three-dimensional coherence. Instead, the vertical organization of disorder, rather than its overall magnitude, plays a decisive role in shaping the *c*-axis electrodynamics by imposing structure onto the interlayer tunneling landscape. This perspective offers a unified interpretation of diverse experimental signatures across cuprate families, highlighting the broader principle that disorder can serve not only as a source of decoherence but also as an organizing agent in strongly anisotropic correlated materials.

### VI. Discussion: Interlayer Coherence, Competing Orders, Predictions, and Limitations of the Vertically Structured Tunneling Framework

The phenomenology developed above places the fragile *c*-axis electrodynamics of cuprates within a broader context in which interlayer coherence is governed not only by intrinsic electronic correlations but also by the spatial organization of disorder. A related conceptual viewpoint was developed in the context of quantum transitions in magnetic molecules, where the evolution of a spin toward its energetically favorable state is governed by a structured landscape of barriers and correlated pathways (Sirenko et al. [45]). Although the underlying microscopic systems differ from cuprates, the emphasis on the *organization* of the energy landscape, rather than on the magnitude of local perturbations, is fully consistent with our finding that vertically correlated disorder, by structuring the tunneling matrix elements $t_\perp(z)$, governs the effective interlayer coherence. In strongly anisotropic systems such as the cuprates, quenched defects, dopants, and strain fields are not merely sources of decoherence. When vertically correlated, they impose a structured modulation onto the interlayer tunneling amplitude $t_\perp(z)$, reshaping the balance among superconductivity, charge order, and magnetic excitations, and modifying the characteristic scales associated with dimensional crossover. In addition to geometric modulation of the tunneling matrix elements, strong scattering environments are known to renormalize quasiparticle lifetimes and redistribute spectral weight in unconventional superconductors, particularly in the unitary scattering limit [46]. While the present work does not rely on a specific microscopic impurity model, such lifetime renormalizations illustrate how correlated disorder can reorganize low-energy excitation spectra without necessarily inducing long-range three-dimensional order. In this sense, vertically structured disorder may act not only as a spatial modulator of $t_\perp(z)$, but also as an extended scattering environment capable of reshaping the effective interlayer energy landscape.

A central example is the relation between vertically structured tunneling and the pair-density-wave (PDW) scenario proposed for La-based cuprates near one-eighth doping. In the PDW picture, orthogonal stripe orientations in adjacent layers frustrate interlayer Josephson coupling, producing robust two-dimensional superconductivity with negligible *c*-axis coherence [13] [15]. The vertically structured tunneling framework does not require a sign-changing superconducting order parameter; however, both perspectives emphasize that misalignment between the electronic environments of neighboring $CuO_2$ planes is sufficient to suppress coherent pair tunneling without reducing in-plane superconductivity. Impurity-pinned CDW textures, known to exhibit topological rigidity under quenched disorder [24], provide a natural microscopic mechanism for such layer-to-layer misalignment.

The interplay between vertically correlated disorder and charge-density-wave correlations offers a further instructive illustration. The CDW order in underdoped cuprates is predominantly two-dimensional [22], yet can acquire a 3D stacking under strong magnetic fields in $YBa_2Cu_3O_{6+\delta}$ [16] [2]. This behavior implies that genuine interlayer CDW coherence requires an external symmetry-breaking field. The vertically structured tunneling picture naturally explains this selectivity: correlated disorder can seed weak vertical correlations among short-range CDW domains, but these correlations remain incoherent unless superconductivity is suppressed. The absence of field-induced 3D stacking in structurally cleaner systems, such as Hg-based cuprates [3], underscores the importance of disorder organization.

These ideas directly connect to dimensional crossover phenomena. Cuprate superconductors frequently exhibit transitions between effectively two-dimensional and three-dimensional regimes in both superconducting fluctuations and *c*-axis charge transport. Phase-fluctuation analyses demonstrate that the superfluid stiffness in underdoped cuprates is intrinsically small and highly anisotropic, so that even modest changes in interlayer coupling can drive pronounced dimensional crossover behavior [47]. Complementary transport studies further show that the *c*-axis response often becomes incoherent or VRH-like even while in-plane transport remains metallic, reflecting the emergence of self-organized, quasi-two-dimensional electronic structures at low doping [12]. Within the present framework, such crossovers arise because vertical segments characterized by different values of $t_\perp$ contribute unequally depending on temperature, magnetic field, oxygen ordering, or annealing history. This behavior echoes long-standing observations that dopants in cuprates reshape not only carrier concentration but the underlying correlated landscape [48].

The vertically structured tunneling perspective also integrates smoothly with existing models of cuprate electrodynamics. Incoherent tunneling descriptions, stripe-based theories, PDW states, two-fluid pictures, and intrinsic Josephson-junction stacks each capture particular facets of *c*-axis behavior. The present framework does not replace these paradigms; rather, it highlights a unifying physical principle: interlayer coherence is extraordinarily sensitive to the spatial arrangement of disorder.

This sensitivity to the morphology and correlated organization of disorder is not unique to cuprates. Closely related behavior, where tunneling pathways are selected and reshaped by anisotropy, particle geometry, and the correlated arrangement of structural units, is well documented in granular oxide systems. In particular, $CrO_2$-based powder composites exhibit (*i*) tunnel magnetoresistance governed by particle-shape anisotropy and its collective alignment, and (*ii*) percolative transport dominated by a small number of high-transparency tunneling paths, leading to strong sample-to-sample variability [49] [36] [50]. Although granular oxides differ fundamentally from layered cuprates, these examples illustrate a broader organizing principle: tunneling-mediated responses in correlated or composite systems are governed not simply by the amount of disorder but by its spatial patterning, especially when only a few effective conduction channels control the transport. Even modest modifications of defect morphology or oxygen distribution can substantially redistribute interlayer magnetic spectral weight. In particular, bilayer magnetic resonance in $YBa_2Cu_3O_7$ is known to consist of odd and even layer modes, whose relative intensities are exceptionally sensitive to disorder. Structural inhomogeneity can redistribute spectral weight between odd/even channels [40]. Consistent with this sensitivity, dilute Zn substitution is likewise known to perturb the magnetic excitation spectrum and redistribute low-energy spin spectral weight [4]. Instances where disorder enhances superconducting coherence [51] further emphasize that disorder can act as an organizing, rather than purely disruptive, element in strongly anisotropic electronic systems.

*Experimentally Testable Implications of the Vertically Structured Tunneling Framework*

The phenomenology developed here does not yield sharp, quantitative predictions; however, it suggests a set of experimentally accessible trends that naturally follow if interlayer tunneling is strongly shaped by vertically correlated disorder:

- *Enhanced multichannel JPR structure in systems with deliberately aligned vertical defects*. If ion irradiation, pressure-driven twin rearrangement, or controlled oxygen-order manipulation introduces partial vertical correlations, the resulting inhomogeneity in $t_\perp(z)$ may manifest as broadened or multi-component Josephson plasma spectra. In contrast, predominantly point-like, uncorrelated disorder is not expected to generate such signatures.

- *Suppression of c-axis anomalies in ultraclean materials.* High-purity Hg-based cuprates, epitaxial films with minimal stacking faults, or carefully annealed single crystals should exhibit single-component JPR spectra and smoother $\rho_c(T)$ curves, consistent with a reduced spread of effective tunneling amplitudes.

- *Tunability of dimensional crossover via vertical correlation length.* Because the vertical organization of dopants and oxygen influences the distribution of local tunneling channels, changes induced by annealing, strain gradients, or interstitial-oxygen redistribution may shift the temperatures at which the system crosses over between 2D- and 3D-dominated fluctuation regimes.

- *Vertically assisted CDW stacking without long-range CDW coherence.* Weak field-induced three-dimensionality of CDW correlations is expected to be most pronounced in materials where disorder already seeds short-range vertical correlations, and least visible in systems with highly uniform oxygen and dopant distributions.

- *Mode-selective sensitivity of bilayer magnetic excitations.* Because odd and even magnetic resonances couple differently to the interlayer structure, preferential alignment of defect or strain columns could differentially perturb the two modes, thereby modifying their relative spectral weights in a predictable manner.

While the vertically structured tunneling framework unifies diverse *c*-axis anomalies, it is essential to recognize its scope and limitations:

- The model does not address the microscopic pairing mechanism or the origin of the pseudogap; its focus is solely on how disorder organization modifies interlayer coherence.

- The description is phenomenological, leaving the microscopic form of $\delta t_\perp(z)$ unspecified. Quantitative predictions for specific materials will require microscopic input from DFT, cluster DMFT, or large-scale lattice simulations.

- Dynamic interlayer phase fluctuations are treated only through their effect on the JPR and dc transport; time-dependent coherence phenomena lie outside the present framework.

- The model assumes weak coupling between vertically defined tunneling segments; if future experiments uncover strongly interacting vertical channels, refinements will be necessary.

Taken together, these considerations indicate that vertically correlated disorder, rather than acting solely as a source of decoherence, can play a constructive role in systems with intrinsically weak interlayer coupling. By imposing spatial structure on the effective tunneling amplitude, disorder can reshape the balance of competing orders, shift dimensional crossover scales, and coordinate nanoscale textures across adjacent layers. This perspective provides a coherent and experimentally accessible framework for interpreting a wide range of *c*-axis anomalies in cuprates, suggesting new avenues for engineering interlayer quantum states through controlled disorder.

## VII. Conclusions

We have introduced a framework in which the *c*-axis electrodynamics of cuprate superconductors arises from the combined effects of intrinsically weak interlayer tunneling and the spatial organization of vertically correlated disorder. In contrast to explanations invoking real charge transfer, electronic reconstruction, or genuine three-dimensional order, the present picture emphasizes that the structure of disorder (not merely its magnitude) governs the effective interlayer coupling in these highly anisotropic materials.

Our analysis shows that interlayer tunneling in cuprates is exceptionally sensitive to nanoscale structural motifs that propagate along the *c*-axis, including interstitial-oxygen staging, twin boundaries, dislocation lines, and vertically repeated impurity configurations. These motifs introduce a layer-dependent modulation of the tunneling amplitude $t_\perp(z)$, naturally generating a multichannel *c*-axis response. This framework accounts for a wide range of experimental observations, including multi-component Josephson plasma resonances, nonmonotonic *c*-axis resistivity, field-enhanced CDW stacking, redistribution of bilayer magnetic spectral weight, and vertically correlated signatures in NMR, μSR, and STM.

By recognizing the key role of correlated disorder, the present work reconciles diverse *c*-axis anomalies across multiple cuprate families without invoking energetically prohibitive interlayer charge redistribution. Instead, we identify fragile interlayer coherence as a sensitive probe of how defects, strain, oxygen mobility, and charge textures self-organize along the *c*-axis. This perspective offers a unifying phenomenology for cuprate electrodynamics, suggesting new pathways for tuning dimensionality, interlayer superconducting coherence, and the competition among electronic orders.

This framework also highlights promising directions for future investigation. Quantitative modeling of $t_\perp(z)$ in realistic disorder landscapes, controlled manipulation of vertical disorder correlations through annealing or strain engineering, and high-resolution probes of layer-resolved electrodynamics can deepen our understanding of interlayer physics in strongly correlated materials. More broadly, the findings underscore how weak interlayer coupling can amplify subtle organizational features of disorder into experimentally significant emergent phenomena — a principle with potential relevance far beyond the cuprates.

### Acknowledgments

The author gratefully acknowledges support from the SARU (Scholars at Risk Ukraine) program, funded by the Carlsberg Foundation, which enabled my research activities under challenging circumstances, as well as the Director of the Mads Clausen Institute, Prof. H.-G. Rubahn, for his continued interest and support.

### References:

[1] D.N. Basov, T. Timusk, Electrodynamics of high-Tc superconductors, Rev. Mod. Phys. 77 (2005) 721–779. https://doi.org/10.1103/RevModPhys.77.721.

[2] J. Chang, E. Blackburn, O. Ivashko, A.T. Holmes, N.B. Christensen, M. Hücker, R. Liang, D.A. Bonn, W.N. Hardy, U. Rütt, M. v. Zimmermann, E.M. Forgan, S.M. Hayden, Magnetic field controlled charge density wave coupling in underdoped YBa2Cu3O6+x, Nat. Commun. 7 (2016) 11494. https://doi.org/10.1038/ncomms11494.

[3] W. Tabis, Y. Li, M. Le Tacon, L. Braicovich, A. Kreyssig, M. Minola, G. Dellea, E. Weschke, M.J. Veit, M. Ramazanoglu, A.I. Goldman, T. Schmitt, G. Ghiringhelli, N. Barišić, M.K. Chan, C.J. Dorow, G. Yu, X. Zhao, B. Keimer, M. Greven, Charge order and its connection with Fermi-liquid charge transport in a pristine high-Tc cuprate, Nat. Commun. 5 (2014) 5875. https://doi.org/10.1038/ncomms6875.

[4] Y. Sidis, P. Bourges, H.F. Fong, B. Keimer, L.P. Regnault, J. Bossy, A. Ivanov, B. Hennion, P.

Gautier-Picard, G. Collin, D.L. Millius, I.A. Aksay, Quantum Impurities and the Neutron Resonance Peak in YBa2Cu3O7: Ni versus Zn, Phys. Rev. Lett. 84 (2000) 5900–5903. https://doi.org/10.1103/PhysRevLett.84.5900.

[5] S. Pailhès, P. Bourges, Y. Sidis, C. Bernhard, B. Keimer, C.T. Lin, J.L. Tallon, Absence of an isotope effect in the magnetic resonance in high-Tc superconductors, Phys. Rev. B 71 (2005) 220507. https://doi.org/10.1103/PhysRevB.71.220507.

[6] E.Y. Beliayev, Y.K. Mishra, I.A. Chicihibaba, I.G. Mirzoiev, V.A. Horielyi, A. V. Terekhov, Emergent Coherence at the Edge of Magnetism: Low-Doped La2-xSrxCuO4+delta Revisited, 2026. https://doi.org/10.48550/arXiv.2602.04452.

[7] J.D. Jorgensen, S. Pei, P. Lightfoor, H. Shi, A.P. Paulikas, B.W. Veal, Time-dependent structural phenomena at room temperature in quenched YBa2Cu3O6.41, Phys. C Supercond. 167 (1990) 571–578. https://doi.org/10.1016/0921-4534(90)90676-6.

[8] S.A. Kivelson, I.P. Bindloss, E. Fradkin, V. Oganesyan, J.M. Tranquada, A. Kapitulnik, C. Howald, How to detect fluctuating stripes in the high-temperature superconductors, Rev. Mod. Phys. 75 (2003) 1201–1241. https://doi.org/10.1103/RevModPhys.75.1201.

[9] Y.-J. Kao, H.-Y. Kee, Anisotropic spin and charge excitations in superconductors: Signature of electronic nematic order, Phys. Rev. B 72 (2005) 024502. https://doi.org/10.1103/PhysRevB.72.024502.

[10] K. Takenaka, K. Mizuhashi, H. Takagi, S. Uchida, Interplane charge transport in YBa2Cu3O7-y: Spin-gap effect on in-plane and out-of-plane resistivity, Phys. Rev. B 50 (1994) 6534–6537. https://doi.org/10.1103/PhysRevB.50.6534.

[11] B.I. Belevtsev, N. V. Dalakova, V.N. Savitsky, A.S. Panfilov, I.S. Braude, A. V. Bondarenko, Magnetoresistive study of the antiferromagnetic–weak ferromagnetic transition in single-crystal La2CuO4+δ, Low Temp. Phys. 30 (2004) 411–416. https://doi.org/10.1063/1.1739162.

[12] Y. Ando, K. Segawa, S. Komiya, A.N. Lavrov, Electrical Resistivity Anisotropy from Self-Organized One Dimensionality in High-Temperature Superconductors, Phys. Rev. Lett. 88 (2002) 137005. https://doi.org/10.1103/PhysRevLett.88.137005.

[13] J.M. Tranquada, B.J. Sternlieb, J.D. Axe, Y. Nakamura, S. Uchida, Evidence for stripe correlations of spins and holes in copper oxide superconductors, Nature 375 (1995) 561–563. https://doi.org/10.1038/375561a0.

[14] Q. Li, M. Hücker, G.D. Gu, A.M. Tsvelik, J.M. Tranquada, Two-Dimensional Superconducting Fluctuations in Stripe-Ordered La1.875Ba0.125CuO4, Phys. Rev. Lett. 99 (2007) 067001. https://doi.org/10.1103/PhysRevLett.99.067001.

[15] E. Berg, E. Fradkin, S.A. Kivelson, J.M. Tranquada, Striped superconductors: how spin, charge and superconducting orders intertwine in the cuprates, New J. Phys. 11 (2009) 115004. https://doi.org/10.1088/1367-2630/11/11/115004.

[16] S. Gerber, H. Jang, H. Nojiri, S. Matsuzawa, H. Yasumura, D.A. Bonn, R. Liang, W.N. Hardy, Z. Islam, A. Mehta, S. Song, M. Sikorski, D. Stefanescu, Y. Feng, S.A. Kivelson, T.P. Devereaux, Z.-X. Shen, C.-C. Kao, W.-S. Lee, D. Zhu, J.-S. Lee, Three-dimensional charge density wave order in YBa2Cu3O6.67 at high magnetic fields, Science (80-. ). 350 (2015) 949–952. https://doi.org/10.1126/science.aac6257.

[17] L.M. Dieng, A.Y. Ignatov, T.A. Tyson, M. Croft, F. Dogan, C.-Y. Kim, J.C. Woicik, J. Grow, Observation of changes in the atomic and electronic structure of single-crystal YBa2Cu3O6.6 accompanying bromination, Phys. Rev. B 66 (2002) 014508.

https://doi.org/10.1103/PhysRevB.66.014508.

[18] M. Tanaka, S. Takebayashi, K. Sawano, S. Kashiwaya, F. Hirayama, M. Koyanagi, Scanning tunneling spectroscopic studies of quench and melt growth (QMG) YBaCuO crystals at 4.2K, Phys. C Supercond. 185–189 (1991) 1909–1910. https://doi.org/10.1016/0921-4534(91)91080-N.

[19] A. Llordés, A. Palau, J. Gázquez, M. Coll, R. Vlad, A. Pomar, J. Arbiol, R. Guzmán, S. Ye, V. Rouco, F. Sandiumenge, S. Ricart, T. Puig, M. Varela, D. Chateigner, J. Vanacken, J. Gutiérrez, V. Moshchalkov, G. Deutscher, C. Magen, X. Obradors, Nanoscale strain-induced pair suppression as a vortex-pinning mechanism in high-temperature superconductors, Nat. Mater. 11 (2012) 329–336. https://doi.org/10.1038/nmat3247.

[20] A.W. Hunt, P.M. Singer, K.R. Thurber, T. Imai, 63Cr NQR Measurement of Stripe Order Parameter in La2-xSrxCuO4, Phys. Rev. Lett. 82 (1999) 4300–4303. https://doi.org/10.1103/PhysRevLett.82.4300.

[21] C.-C. Hsu, B.-C. Huang, M. Schnedler, M.-Y. Lai, Y.-L. Wang, R.E. Dunin-Borkowski, C.-S. Chang, T.-K. Lee, P. Ebert, Y.-P. Chiu, Atomically-resolved interlayer charge ordering and its interplay with superconductivity in YBa2Cu3O6.81, Nat. Commun. 12 (2021) 3893. https://doi.org/10.1038/s41467-021-24003-0.

[22] G. Ghiringhelli, M. Le Tacon, M. Minola, S. Blanco-Canosa, C. Mazzoli, N.B. Brookes, G.M. De Luca, A. Frano, D.G. Hawthorn, F. He, T. Loew, M.M. Sala, D.C. Peets, M. Salluzzo, E. Schierle, R. Sutarto, G.A. Sawatzky, E. Weschke, B. Keimer, L. Braicovich, Long-Range Incommensurate Charge Fluctuations in (Y,Nd)Ba2Cu3O6+x, Science (80-. ). 337 (2012) 821–825. https://doi.org/10.1126/science.1223532.

[23] J. Chang, E. Blackburn, A.T. Holmes, N.B. Christensen, J. Larsen, J. Mesot, R. Liang, D.A. Bonn, W.N. Hardy, A. Watenphul, M. v. Zimmermann, E.M. Forgan, S.M. Hayden, Direct observation of competition between superconductivity and charge density wave order in YBa2Cu3O6.67, Nat. Phys. 8 (2012) 871–876. https://doi.org/10.1038/nphys2456.

[24] M. Kang, J. Pelliciari, A. Frano, N. Breznay, E. Schierle, E. Weschke, R. Sutarto, F. He, P. Shafer, E. Arenholz, M. Chen, K. Zhang, A. Ruiz, Z. Hao, S. Lewin, J. Analytis, Y. Krockenberger, H. Yamamoto, T. Das, R. Comin, Evolution of charge order topology across a magnetic phase transition in cuprate superconductors, Nat. Phys. 15 (2019) 335–340. https://doi.org/10.1038/s41567-018-0401-8.

[25] A.J. Sievers, S. Takeno, Intrinsic Localized Modes in Anharmonic Crystals, Phys. Rev. Lett. 61 (1988) 970–973. https://doi.org/10.1103/PhysRevLett.61.970.

[26] S. Chakravarty, A. Sudbø, P.W. Anderson, S. Strong, Interlayer Tunneling and Gap Anisotropy in High-Temperature Superconductors, Science (80-. ). 261 (1993) 337–340. https://doi.org/10.1126/science.261.5119.337.

[27] K. Tamasaku, Y. Nakamura, S. Uchida, Charge dynamics across the CuO2 planes in La2-xSrxCuO4, Phys. Rev. Lett. 69 (1992) 1455–1458. https://doi.org/10.1103/PhysRevLett.69.1455.

[28] S. Uchida, T. Ido, H. Takagi, T. Arima, Y. Tokura, S. Tajima, Optical spectra of La2-xSrxCuO4: Effect of carrier doping on the electronic structure of the CuO2 plane, Phys. Rev. B 43 (1991) 7942–7954. https://doi.org/10.1103/PhysRevB.43.7942.

[29] A. Gumann, T. Dahm, N. Schopohl, Microscopic theory of superconductor-constriction-superconductor Josephson junctions in a magnetic field, Phys. Rev. B 76 (2007) 064529. https://doi.org/10.1103/PhysRevB.76.064529.

[30] A.L. Solovjov, K. Rogacki, N. V. Shytov, E. V. Petrenko, L. V. Bludova, A. Chroneos, R. V. Vovk, Influence of strong electron irradiation on fluctuation conductivity and pseudogap in YBa2Cu3O7-d single crystals, Phys. Rev. B 111 (2025) 174508. https://doi.org/10.1103/PhysRevB.111.174508.

[31] E. Dagotto, Complexity in Strongly Correlated Electronic Systems, Science (80-. ). 309 (2005) 257–262. https://doi.org/10.1126/science.1107559.

[32] S. V. Dordevic, S. Komiya, Y. Ando, D.N. Basov, Josephson Plasmon and Inhomogeneous Superconducting State in La2-xSrxCuO4, Phys. Rev. Lett. 91 (2003) 167401. https://doi.org/10.1103/PhysRevLett.91.167401.

[33] J.D. Jorgensen, B. Dabrowski, S. Pei, D.G. Hinks, L. Soderholm, B. Morosin, J.E. Schirber, E.L. Venturini, D.S. Ginley, Superconducting phase of La2CuO4+d: A superconducting composition resulting from phase separation, Phys. Rev. B 38 (1988) 11337–11345. https://doi.org/10.1103/PhysRevB.38.11337.

[34] N. Kameda, M. Tokunaga, T. Tamegai, M. Konczykowski, S. Okayasu, Josephson plasma resonance in Bi2Sr2CaCu2O8+y with spatially dependent interlayer phase coherence, Phys. Rev. B 69 (2004) 180502. https://doi.org/10.1103/PhysRevB.69.180502.

[35] J. Fiore, N. Sellati, F. Gabriele, C. Castellani, G. Seibold, M. Udina, L. Benfatto, Investigating Josephson plasmons in layered cuprates via nonlinear terahertz spectroscopy, Phys. Rev. B 110 (2024) L060504. https://doi.org/10.1103/PhysRevB.110.L060504.

[36] B.I. Belevtsev, N. V. Dalakova, M.G. Osmolowsky, E.Y. Beliayev, A.A. Selutin, Y.A. Kolesnichenko, Percolation effects in the conductivity and magnetoresistance of compacted chromium dioxide powder, Bull. Russ. Acad. Sci. Phys. 74 (2010) 1062–1065. https://doi.org/10.3103/S1062873810080071.

[37] B.I. Belevtsev, E.Y. Belyaev, Y.F. Komnik, E.Y. Kopeichenko, Transition from strong to weak electron localization in a percolating gold film under the influence of an electric field, Low Temp. Phys. 23 (1997) 724–732. https://doi.org/10.1063/1.593369.

[38] B.I. Belevtsev, Y.F. Komnik, E.Y. Belyaev, Electron phase relaxation in disordered gold films irradiated with Ar ions, Low Temp. Phys. 21 (1995) 646–654. https://doi.org/10.1063/10.0033861.

[39] T. Wu, H. Mayaffre, S. Krämer, M. Horvatić, C. Berthier, W.N. Hardy, R. Liang, D.A. Bonn, M.-H. Julien, Incipient charge order observed by NMR in the normal state of YBa2Cu3Oy, Nat. Commun. 6 (2015) 6438. https://doi.org/10.1038/ncomms7438.

[40] H.F. Fong, B. Keimer, D. Reznik, D.L. Milius, I.A. Aksay, Polarized and unpolarized neutron-scattering study of the dynamical spin susceptibility of YBa2Cu3O7, Phys. Rev. B 54 (1996) 6708–6720. https://doi.org/10.1103/PhysRevB.54.6708.

[41] A.T. Savici, Y. Fudamoto, I.M. Gat, T. Ito, M.I. Larkin, Y.J. Uemura, G.M. Luke, K.M. Kojima, Y.S. Lee, M.A. Kastner, R.J. Birgeneau, K. Yamada, Muon spin relaxation studies of incommensurate magnetism and superconductivity in stage-4 La2CuO4.11 and La1.88Sr0.12CuO4, Phys. Rev. B 66 (2002) 014524. https://doi.org/10.1103/PhysRevB.66.014524.

[42] S.H. Pan, J.P. O'Neal, R.L. Badzey, C. Chamon, H. Ding, J.R. Engelbrecht, Z. Wang, H. Eisaki, S. Uchida, A.K. Gupta, K.-W. Ng, E.W. Hudson, K.M. Lang, J.C. Davis, Microscopic electronic inhomogeneity in the high-Tc superconductor Bi2Sr2CaCu2O8+x, Nature 413 (2001) 282–285. https://doi.org/10.1038/35095012.

[43] Y. Kohsaka, C. Taylor, K. Fujita, A. Schmidt, C. Lupien, T. Hanaguri, M. Azuma, M. Takano, H. Eisaki, H. Takagi, S. Uchida, J.C. Davis, An Intrinsic Bond-Centered Electronic Glass with Unidirectional Domains in Underdoped Cuprates, Science (80-. ). 315 (2007) 1380–1385. https://doi.org/10.1126/science.1138584.

[44] J.A. Slezak, J. Lee, M. Wang, K. McElroy, K. Fujita, B.M. Andersen, P.J. Hirschfeld, H. Eisaki, S. Uchida, J.C. Davis, Imaging the impact on cuprate superconductivity of varying the interatomic distances within individual crystal unit cells, Proc. Natl. Acad. Sci. 105 (2008) 3203–3208. https://doi.org/10.1073/pnas.0706795105.

[45] V. Sirenko, F.U. Bartolomé, J. Bartolomé, The paradigm of magnetic molecule in quantum matter: Slow molecular spin relaxation, Low Temp. Phys. 50 (2024) 431–445. https://doi.org/10.1063/10.0026056.

[46] P. Contreras, D. Osorio, E.Y. Beliayev, Dressed behavior of the quasiparticles lifetime in the unitary limit of two unconventional superconductors, Low Temp. Phys. 48 (2022) 187–192. https://doi.org/10.1063/10.0009535.

[47] V.J. Emery, S.A. Kivelson, Importance of phase fluctuations in superconductors with small superfluid density, Nature 374 (1995) 434–437. https://doi.org/10.1038/374434a0.

[48] H. Alloul, What do we learn from impurities and disorder in high-Tc cuprates?, Front. Phys. 12 (2024). https://doi.org/10.3389/fphy.2024.1406242.

[49] N. V. Dalakova, E.Y. Beliayev, O.M. Osmolovskaya, M.G. Osmolovsky, V.A. Gorelyy, Tunnel magnetoresistance of compacted CrO2 powders with particle shape anisotropy, Bull. Russ. Acad. Sci. Phys. 79 (2015) 789–793. https://doi.org/10.3103/S1062873815060064.

[50] E.Y. Beliayev, V.A. Horielyi, Y.A. Kolesnichenko, Magnetotransport properties of CrO2 powder composites (Review article), Low Temp. Phys. 47 (2021) 355–377. https://doi.org/10.1063/10.0004228.

[51] M. Leroux, V. Mishra, J.P.C. Ruff, H. Claus, M.P. Smylie, C. Opagiste, P. Rodière, A. Kayani, G.D. Gu, J.M. Tranquada, W.-K. Kwok, Z. Islam, U. Welp, Disorder raises the critical temperature of a cuprate superconductor, Proc. Natl. Acad. Sci. 116 (2019) 10691–10697. https://doi.org/10.1073/pnas.1817134116.